\documentclass
[floatfix,superscriptaddress,secnumarabic,amssymb,amsmath,nobibnotes,aps,prd,showkeys,showpacs,nofootinbib,onecolumn,notitlepage,12pt]{revtex4-2}%
\usepackage{setspace}
\usepackage{color}
\usepackage{amsmath}
\usepackage{amsfonts}
\usepackage{verbatim}
\usepackage{amssymb}
\usepackage{graphicx,bm}
\usepackage{graphicx}
\usepackage{amsmath}
\usepackage{amssymb}
\usepackage{amssymb}
\usepackage{graphicx,bm}
\usepackage{graphicx}
\usepackage[caption=false]{subfig}%
\providecommand{\U}[1]{\protect\rule{.1in}{.1in}}

\newcommand{\be}{\begin{equation}}
\newcommand{\ee}{\end{equation}}

\newcommand{\mincir}{\raise
-3.truept\hbox{\rlap{\hbox{$\sim$}}\raise4.truept\hbox{$<$}\ }}
\newcommand{\magcir}{\raise
-3.truept\hbox{\rlap{\hbox{$\sim$}}\raise4.truept\hbox{$>$}\ }}

\ifx\pdfoutput\relax\let\pdfoutput=\undefined\fi
\newcount\msipdfoutput
\ifx\pdfoutput\undefined\else
\ifcase\pdfoutput\else
\ifx\paperwidth\undefined\else
\ifdim\paperheight=0pt\relax\else\pdfpageheight\paperheight\fi
\ifdim\paperwidth=0pt\relax\else\pdfpagewidth\paperwidth\fi
\fi\fi\fi
\begin{document}
\title{Exploring Quantum Cosmology within the Framework of Teleparallel $f(T)$-gravity}
\author{N. Dimakis}
\email{nikolaos.dimakis@ufrontera.cl}
\affiliation{Departamento de Ciencias F\'isicas, Universidad de la Frontera, Casilla 54-D, 4811186, Temuco, Chile}
\author{A. Paliathanasis}
\email{anpaliat@phys.uoa.gr}
\affiliation{Institute of Systems Science, Durban University of Technology, Durban 4000,
South Africa}
\affiliation{Departamento de Matem\'{a}ticas, Universidad Cat\'{o}lica del Norte, Avda.
Angamos 0610, Casilla 1280 Antofagasta, Chile}
\author{T. Christodoulakis}
\email{tchris@phys.uoa.gr}
\affiliation{Nuclear and Particle Physics section, Physics Department, University of
Athens, 15771 Athens, Greece}

\begin{abstract}
We investigate quantum cosmology in teleparallel $f(T)$-gravity. We delve
extensively into the minisuperspace description within the context of
teleparallelism. The $f(T)$-theory constitutes a second-order theory of
gravity, whose cosmological counterpart is delineated by a degenerate point-like Lagrangian. To formulate
the Hamiltonian function encompassing all constraints and degrees of freedom
inherent to $f(T)$ cosmology, we employ the Dirac-Bergmann algorithm.
Subsequently, we determine the wave function of the universe and introduce a ``probabilistic'' interpretation. We perform comparisons to some classical solutions to see to what extent the quantum approach can cure classical singularities.

\end{abstract}
\maketitle

\section{Introduction}

Classic cosmology describes the large-scale structure and the dynamics of the
universe. Yet, in order to investigate the quantum phenomena that governed the
universe's nascent phases, the field of quantum cosmology comes into play.
The quantum cosmology we know is based on the treatment of reduced versions of classical gravitational systems, and the absence of a comprehensive theory of quantum gravity leaves
room for theoretical discourse regarding its interpretation \cite{rev1,rev2,rev2a,rev2b,rev3}. Nevertheless, quantum cosmology is used in order to understand
the initial state of the universe and the evolution of the quantum
fluctuations in the inflationary mechanism \cite{rev4,rev5}. Novel attempts in the
mathematical construction of a probability density for the
possible states of our cosmos, based on the wave function of the universe, have been presented \cite{sw0,sw0a,sw1,sw1a,sw1b}. It has been claimed that  the quantum
fluctuations can prevent the initial singularity at the very early period of
the universe \cite{sw2,sw3,sw4,sw5,sw6}. For more in-depth exploration and
examination of (the various realizations of) quantum cosmology, we direct the reader to
\cite{sm0,sm1,sm1a,sm1b,sm1c,sm1d,sm2,sm3,sm4,sm5,sm6} and references therein.

Meanwhile, at the classical level, and in order to explain the acceleration phases of the universe, cosmologists have
proposed various gravitational models which are classified in two-large
families. The first family encompasses the modified, or alternative, theories
of gravity, wherein new geometric invariants are introduced, to alter the
Einstein-Hilbert Action of General Relativity (GR), which drive the dynamics
and provide the necessary acceleration \cite{md1,md2,md2b,md3,md4}. In contrast, the second
family consists of models where the gravitational theory remains that of GR,
and the cosmic acceleration is explained by introducing specific matter
components into Einstein's field equations \cite{ss1,ss2,ss3,ss5,ss6,ss7,ss8}.

In this study we investigate quantum cosmology in the framework of
teleparallel $f\left(  T\right)  $-gravity. It is a special class of a set of modified
gravitational theories comprised by: the metric $f(R)$-theory \cite{fr},
teleparallel $f(T)$-theory \cite{Ferraro}, and symmetric teleparallel
$f(Q)$-theory \cite{fq}. In these gravitational frameworks, the gravitational
Lagrangian assumes a functional form denoted as $f$, utilizing one of the
fundamental geometric scalars: the Ricci scalar $R$ for the Levi-Civita
connection, the torsion scalar $T$ for the Weitzenb{\"{o}}ck connection
\cite{Weitzenb23}, and the non-metricity scalar $Q$ for a symmetric and flat
connection. When the function $f$ takes on a linear expression, these three
gravitational models seamlessly converge to General Relativity (GR), the
Teleparallel Equivalent of GR (TEGR) \cite{ngr1}, and Symmetric Teleparallel
GR (STGR) \cite{ngr2} respectively \cite{ngr2b}. Consequently, these three theories are
equivalent in their linear Lagrangian limit. This equivalence can be
intuitively understood noting the fact that the three scalars, namely $R$,
$T$, and $Q$, differ by a total derivative term, which can be omitted, as far
as the process of variation is concerned \cite{fd1,fd2}.

Despite some existing research on this particular topic \cite{ds2,ds3,ds1}, here we follow a different and more extensive procedure. We use the Dirac-Bergmann algorithm \cite{DB1,DB2}, in order to reveal the constraints of the system and formally obtain the Hamiltonian function to be used in the quantization. In works like \cite{ds2,ds3} the Dirac-Bergmann algorithm is ignored altogether and, as a result, there is no distinction among first and second class constraints, thus, leading to a compromised quantization of a Hamiltonian which contains non-physical degrees of freedom. This particular issue has also been recently encountered in the exploration of quantum cosmology within $f(Q)$-gravity \cite{Cap}. However, in \cite{Cap}, an additional error arises due to an incorrect assumption of gravitational field equations in the presence of nonzero spatial curvature for the background spacetime. In \cite{ds1}, on the other hand, the Dirac-Bergmann algorithm is followed correctly for $f(T)$ cosmology and a spatially flat Friedmann--Lema\^{\i}tre--Robertson--Walker (FLRW) space-time in the presence of a cosmological constant term. The authors derive the constraints of the theory and choose an approach of promoting, with appropriate reductions, the second class relations to first class, which they later use to quantize the relative system. Here, we adopt a more traditional approach, and for a FLRW space-time with a perfect fluid matter content, we use the second class constraints to introduce Dirac brackets \cite{Diracbook} with respect to which we proceed to the canonical quantization.


Quantum cosmology in symmetric teleparallel $f\left(  Q\right)  $-gravity was
the subject of our study in \cite{QuantfQus} where we presented an approach for
constructing a Hamiltonian function in $f(Q)$-theory that encompasses all
first and second class constraints. In many aspects the process in the $f\left( T\right)$ theory is similar, containing however some key difference mostly when the spatial curvature is nonzero. The study of quantum cosmology
within $f(R)$-gravity was conducted in \cite{DimfR,AndfR}. However, the
dynamics in $f(R)$-theory differ from that of $f\left(  T\right)  $ and
$f\left(  Q\right)  $ theories due to the association of the theory's degrees
of freedom with a scalar field. The constraints in this theory align with
those of scalar-tensor theory \cite{AndfR}. The structure of the paper is
outlined below.

In Section \ref{sec2}, we introduce the geometric trinity of gravity and
discuss the primary differences in the corresponding gravitational theories.
The minisuperspace description for the cosmological field equations is
discussed in Section \ref{sec3}, where we review previous studies on the
subject and examine their weaknesses. Section \ref{sec4} presents the key
findings of this work, where we introduce a complete quantization for the
minisuperspace process in teleparallel $f(T)$-gravity within a FLRW geometry
with a matter source. The presence of the matter source in $f(T)$-cosmology is
crucial; otherwise, the unique classical solution of the field equations is
that of vacuum GR \cite{bl}. Finally, in Section \ref{sec5}, we provide a
summary of our results.

\section{The geometrical trinity and its modifications}

\label{sec2}

In the attempt to use geometry to formulate a gravitation theory, there can be
employed one or more of three fundamental geometric scalars. Let us start from the
definition of a general connection whose components can be split into three
parts
\begin{equation}
\Gamma_{\phantom{\lambda}\mu\nu}^{\lambda}=\tilde{\Gamma}%
_{\phantom{\lambda}\mu\nu}^{\lambda}+K_{\phantom{\lambda}\mu\nu}^{\lambda
}+L_{\phantom{\lambda}\mu\nu}^{\lambda}. \label{generalcon}%
\end{equation}
The first
\begin{equation}
\tilde{\Gamma}_{\phantom{\lambda}\mu\nu}^{\lambda}=\frac{1}{2}g^{\kappa
\lambda}\left(  \frac{\partial g_{\kappa\nu}}{\partial x^{\mu}}+\frac{\partial
g_{\mu\kappa}}{\partial x^{\nu}}-\frac{\partial g_{\mu\nu}}{\partial
x^{\kappa}}\right)  ,
\end{equation}
given a spacetime metric $g_{\mu\nu}$, is made up by the Christoffel symbols. The second, consists of the contorsion tensor, which is defined as
\begin{equation}
K_{\phantom{\lambda}\mu\nu}^{\lambda}=\frac{1}{2}\left(  T_{\mu
\phantom{\lambda}\nu}^{\phantom{\mu}\lambda}+T_{\nu\phantom{\lambda}\mu
}^{\phantom{\mu}\lambda}-T_{\phantom{\lambda}\mu\nu}^{\lambda}\right)  ,
\label{contor}%
\end{equation}
where
\begin{equation}
T_{\phantom{\lambda}\mu\nu}^{\lambda}=\Gamma_{\phantom{\lambda}\nu\mu
}^{\lambda}-\Gamma_{\phantom{\lambda}\mu\nu}^{\lambda} \label{tortensor}%
\end{equation}
is the torsion. Finally, the last part
\begin{equation}
L_{\phantom{\lambda}\mu\nu}^{\lambda}=\frac{1}{2}g^{\lambda\kappa}\left(
Q_{\mu\nu\kappa}+Q_{\nu\mu\kappa}+Q_{\kappa\mu\nu}\right)
\end{equation}
is the disformation tensor, which emerges due to the presence of nonmetricity
\begin{equation}
Q_{\lambda\mu\nu}=\nabla_{\lambda}g_{\mu\nu},
\end{equation}
where $\nabla_{\lambda}$ is the covariant derivative with respect to the
connection \eqref{generalcon}.

In the case of the usual (pseudo-)Riemannian geometry, one considers a
torsionless, $T_{\phantom{\lambda}\mu\nu}^{\lambda}=0$, connection with zero
nonmetricity, $Q_{\lambda\mu\nu}=0$, which leads the connection
\eqref{generalcon} to be given just in terms of the Christoffel symbols
$\Gamma_{\phantom{\lambda}\mu\nu}^{\lambda}=\tilde{\Gamma}%
_{\phantom{\lambda}\mu\nu}^{\lambda}$. Subsequently, gravitational effects are
assigned to the Riemannian curvature and its scalar $R$. The action of General
Relativity is linear in the Ricci scalar $R$ and its most famous
generalizations are in terms of $f(R)$ modifications of the Lagrangian
density. Similarly, one may adopt a torsionless and flat geometry, thus
attributing the gravitational phenomena to the nonmetricity $Q_{\lambda\mu\nu
}$ and its corresponding nonmetricity scalar $Q$. In the same manner, a flat
geometry (under the assumption of a Weitzenb\"{o}ck connection) can be chosen
together with $Q_{\lambda\mu\nu}=0$, leaving this time only the torsion
$T_{\phantom{\lambda}\mu\nu}^{\lambda}\neq0$ and its corresponding scalar to
account for gravity. In both of these two last cases,
modifications in the form of $f(Q)$ or $f(T)$ theories can be taken respectively.
These are the simplest geometrical configurations that can be considered, where
in each case, only one of the three basic geometric scalars is nonzero. The
linearized version of this set of theories has been deemed as the geometrical
trinity of gravity \cite{Heistrin1,Koitrin}.

Here, we are going to consider the case of $f(T)$ theories constructed with
the help of the torsion scalar
\begin{equation}
T=S_{\rho}^{\phantom{\rho}\mu\nu}T_{\phantom{\rho}\mu\nu}^{\rho},
\label{Tscalargen}%
\end{equation}
where
\begin{equation}
S^{\rho\mu\nu}=\frac{1}{2}\left(  K^{\mu\nu\rho}-g^{\rho\nu}%
T_{\phantom{\lambda\mu}\lambda}^{\lambda\mu}+g^{\rho\mu}%
T_{\phantom{\lambda\mu}\lambda}^{\lambda\nu}\right)  \label{Stensor}%
\end{equation}
is called the superpotential. As is well known, theories whose Lagrangian
density is linear to the scalar $T$ defined above, i.e. $f(T)\sim T$, are
dynamically equivalent to Einstein's General Relativity. The linear in $T$
theory is referred as the teleparallel equivalent of General Relativity.
Similarly, one can define a nonmetricity scalar $Q$, in such a manner, so as
by taking a theory linear to it the symmetric teleparallel of General
Relativity is obtained \cite{Heistrin2}.

Although $f(R)$, $f(Q)$ and $f(T)$ have a lot in common in the way the
theories are constructed, their dynamical behavior is quite different when we
deviate from the linear expressions with respect to the fundamental scalars of
each case. The $f(R)$ gravity leads to higher order equations, which
effectively correspond to an inclusion of an extra degree of freedom since
the theory is dynamically equivalent to GR with an appropriate addition of a
scalar field. In the cases of $f(Q)$ and $f(T)$ gravity we have second order
equations and, when the connection does not play a dynamical role, it leads
to important implications for the procedure which we need to follow in order to perform
a minisuperspace quantization. We shall refer to this point more extensively
in the subsequent sections.

\section{Minisuperspace Lagrangians}

\label{sec3}

Let us begin by writing the FLRW line element:
\begin{equation}
ds^{2}=-N(t)dt^{2}+a(t)^{2}\left[  d\chi^{2}+\chi^{2}\mathrm{sinc}^{2}%
(\sqrt{k}\chi)\left(  d\theta^{2}+\sin^{2}\theta d\phi^{2}\right)  \right]  ,
\label{lineel}%
\end{equation}
where $\mathrm{sinc}^{2}(\sqrt{k}\chi)=\left[  \frac{\sin(\sqrt{k}\chi)}{\sqrt{k}\chi
}\right]  ^{2}$ is the square of the sine cardinal function. The above line
element can describe at the same time the spatially curved $k=\pm1$ and the
flat $k=0$ cases, since $\underset{k\rightarrow0}{\lim}\;\left(  \frac
{\sin(\sqrt{k}\chi)}{\sqrt{k}\chi}\right)  =1=\mathrm{sinc}(0)$. In these coordinates,
$\chi$ is considered as a radial angle, while the scale factor $a(t)$ has
units of distance.

We can of course express the line element in the form
\begin{equation}
ds^{2}=\eta_{AB}e_{\phantom{A}\mu}^{A}e_{\phantom{A}\nu}^{B}dx^{\mu}dx^{\nu
},\label{lineelvier}%
\end{equation}
where $\eta_{AB}=\mathrm{diag}(-1,1,1,1)$ is the Minkowski metric, and the dual
vierbein is given by {\tiny
\begin{equation}
e_{\phantom{A}\mu}^{A}=%
\begin{pmatrix}
N & 0 & 0 & 0\\
0 & -a\cos\theta & \frac{a}{\sqrt{k}}\left(  \sin(\sqrt{k}\chi)\cos(\sqrt
{k}\chi)\sin\theta\right)   & -\frac{a}{\sqrt{k}}\sin^{2}(\sqrt{k}\chi
)\sin^{2}\theta\\
0 & a\sin\theta\cos\phi & \frac{a}{\sqrt{k}}\sin(\sqrt{k}\chi)\left(
\cos(\sqrt{k}\chi)\cos\theta\cos\phi-\sin(\sqrt{k}\chi)\sin\phi\right)   &
-\frac{a}{\sqrt{k}}\sin(\sqrt{k}\chi)\left(  \sin(\sqrt{k}\chi)\cos\theta
\cos\phi+\cos(\sqrt{k}\chi)\sin\phi\right)  \sin\theta\\
0 & -a\sin\theta\sin\phi & -\frac{a}{\sqrt{k}}\sin(\sqrt{k}\chi)\left(
\cos(\sqrt{k}\chi)\cos\theta\sin\phi+\sin(\sqrt{k}\chi)\cos\phi\right)   &
\frac{a}{\sqrt{k}}\sin(\sqrt{k}\chi)\left(  \sin(\sqrt{k}\chi)\cos\theta
\sin\phi-\cos(\sqrt{k}\chi)\cos\phi\right)  \sin\theta
\end{pmatrix}
\mathbf{.}\label{vierbein}%
\end{equation}
}Note that the above choice is not unique, e.g. one may take the obvious
diagonal vierbein which can be used in \eqref{lineelvier} to reproduce
\eqref{lineel}. However, the utilization of \eqref{vierbein} is crucial in
order to derive a valid minisuperspace Lagrangian. It can be seen that the
latter is a generalization of the diagonal vierbein under the action of
rotation matrices \cite{Ferraro,Coley}.

If we adopt the Weitzenb\"{o}ck connection we get
\begin{equation}
\Gamma_{\phantom{\lambda}\mu\nu}^{\lambda}=e_{A}^{\phantom{A}\lambda}%
\partial_{\nu}e_{\phantom{A}\mu}^{A}, \label{gammacon}%
\end{equation}
which together with \eqref{vierbein} leads to the usual expression for the
torsion scalar in FLRW geometry
\begin{equation}
T=6\left(  \frac{\dot{a}}{Na}\right)  ^{2}-\frac{6k}{a^{2}}. \label{Tscalar}%
\end{equation}

The general action in the case of the $f(T)$ theories of gravity is
\begin{equation}
\mathcal{A}=\mathcal{A}_{grav}+\mathcal{A}_{m}=\frac{1}{2\kappa}%
\int\!\!ef(T)d^{4}x+\mathcal{A}_{m},\label{act}%
\end{equation}
where $e=\det(e_{\phantom{A}\mu}^{A})$ and $\mathcal{A}_{m}$ is the part of the action
which refers to the matter content. The field equations which are obtained by
variation with respect to $e_{\phantom{A}\nu}^{A}$ are
\begin{equation}
\frac{2}{e}\partial_{\mu}\left(  ee_{\phantom{\rho}A}^{\rho}S_{\rho
}^{\phantom{\rho}\mu\nu}f^{\prime}(T)\right)  +2f^{\prime}%
S_{\phantom{\sigma}\mu\rho}^{\sigma}e_{\phantom{\rho}A}^{\rho}S_{\sigma
}^{\phantom{\sigma}\mu\nu}-\frac{1}{2}e_{A}^{\phantom{A}\nu}f(T)+\kappa
e_{A}^{\phantom{A}\rho}\mathcal{T}_{\phantom{\nu}\rho}^{\nu}=0,\label{feq}%
\end{equation}
with $\mathcal{T}_{\phantom{\mu}\nu}^{\mu}$ expressing the energy momentum tensor for
the matter. In our case we are going to consider a perfect fluid of
energy density $\rho(t)$ and pressure $p(t)$ with
\begin{equation}
\mathcal{T}_{\phantom{\mu}\nu}^{\mu}=(\rho+p)v_{\mu}v_{\nu}+pg_{\mu\nu},
\end{equation}
where $v^{\mu}$ is the comoving velocity satisfying $v^{\mu}v_{\mu}=-1$. For
the FLRW metric we consider here, we finally get the well known expression $\mathcal{T}_{\phantom{\mu}\nu}^{\mu
}=\mathrm{diag}(-\rho,p,p,p)$.

The object of writing a minisuperspace Lagrangian is to start from
\eqref{act}, use the ansatz we made for the spacetime at hand and then, obtain a
reduced Lagrangian of finite degrees of freedom, which correctly reproduces
the result of \eqref{feq} for the given spacetime. For example, let us first
take the gravitational part of \eqref{act} and substitute in it
\begin{equation}
e=\det(e_{\phantom{A}\mu}^{A})=Na^{3}\mathrm{sinc}^{2}(\sqrt{k}\chi)\sin\theta
\end{equation}
which is obtained by the use of \eqref{vierbein}. The gravitational part of
the action then factorizes into two sub-parts
\[
\mathcal{A}_{grav}=\frac{1}{2\kappa}\int\!\!ef(T)d^{4}x=\overbrace{\left[
\int\!\!\left(  \chi^{2}\mathrm{sinc}^{2}(\sqrt{k}\chi)\sin\theta\right)
d^{3}x\right]  }^{\text{spatial part}}\overbrace{\frac{1}{2\kappa}%
\int\!\!Na^{3}f(T)dt}^{t-\text{dependent dynamical part}}.
\]
One part has a spatial dependence $(\chi,\theta,\phi)$, while the other is
dynamical, i.e. it depends purely on the time variable $t$. For the
construction of the minisuperspace gravitational Lagrangian $L_{grav}$, we
keep the dynamical $t$-dependent part of the above relation. The spatial part
can be discarded under assumptions of performing an integration in a fixed
volume in space, which would turn it into a constant number.

In the dynamical part of the action we also want to introduce the information
that the torsion scalar is related to the scale factor through Eq.
\eqref{Tscalar}. We thus complete the gravitation part of the reduced
emanating Lagrangian, by adding the relation which gives the torsion scalar
together with a Lagrange multiplier $\lambda(t)$. In other words, we write
\begin{equation}
L_{grav}=\frac{1}{2\kappa}Na^{3}f(T)+\lambda\left[  T-\left(  6\left(
\frac{\dot{a}}{Na}\right)  ^{2}-\frac{6k}{a^{2}}\right)  \right]  .
\label{Lgrav}%
\end{equation}
The variation of $L_{grav}$ with respect to $T$ yields the value of the
multiplier
\begin{equation}
\frac{\partial L_{grav}}{\partial T}=0\Rightarrow\lambda=-\frac{1}{2\kappa
}Na^{3}f^{\prime}(T),
\end{equation}
which we can replace back into \eqref{Lgrav} to finally obtain
\begin{equation}
L_{grav}=\frac{1}{2\kappa}\left[  \frac{6a\dot{a}^{2}f^{\prime}(T)}{N}%
+Na^{3}\left(  f(T)-Tf^{\prime}(T)\right)  -6kNaf^{\prime}(T)\right]  .
\label{Lgrav2}%
\end{equation}

Now, the matter part of the action \eqref{act} for a perfect fluid is
\[
\mathcal{A}_{m}=\int\!\!e\rho d^{4}x.
\]
In order to variate it we need to provide an equation of state and take into
account the continuity equation $T_{\phantom{\mu}\nu;\mu}^{\mu}=0$, where the
semicolon stands for the covariant derivative with respect to the Levi-Civita
connection for FLRW.

When $p=w\rho$, the $T_{\phantom{\mu}\nu;\mu}^{\mu}=0$ yields
\begin{equation}
\mathcal{T}_{\phantom{\mu}\nu;\mu}^{\mu}=0\Rightarrow\dot{\rho}+\frac{\dot{a}%
}{a}\left(  p+\rho\right)  \Rightarrow\rho=\rho_{0}a^{-3(1+w)}, \label{conteq}%
\end{equation}
where $\rho_{0}$ is a constant of integration. By concentrating again on the
$t$-dependent part of $\mathcal{A}_{m}$, we can write the minisuperspace contribution
due to matter as
\begin{equation}
L_{m}=N\rho_{0}a^{-3w}.
\end{equation}

By adding the two contributions, we finally write the total Lagrangian as
$L=L_{grav}+L_{m}$ which results in the expression
\begin{equation}
L=\frac{1}{2\kappa}\left[  \frac{6a\dot{a}^{2}f^{\prime}(T)}{N}+N\left(
a^{3}\left(  f(T)-Tf^{\prime}(T)\right)  -6kaf^{\prime}(T)\right)  \right]
+N\rho_{0}a^{-3w}. \label{Lag}%
\end{equation}
Variation of the above Lagrangian with respect to $N$ and $a$ produces
equations equivalent to those of \eqref{feq} for the assumed space-time.
Variation with respect to $T$ gives rise to the result of definition
\eqref{Tscalar}. Thus, Lagrangian \eqref{Lag} is a valid minisuperspace
Lagrangian which generates correctly the classical dynamics. The
Euler-Lagrange equations can be written equivalently as the set:
\begin{subequations}
\label{euleqs}%
\begin{align}
&  \frac{6\dot{a}^{2}f^{\prime}(T)}{a^{2}N^{2}}-\kappa\rho_{0}a^{-3(w+1)}%
-\frac{1}{2}f(T)=0\\
&  2\frac{d}{dt}\left(  \frac{\dot{a}f^{\prime}(T)}{aN^{2}}\right)
+\frac{2\dot{a}\dot{N}f^{\prime}(T)}{aN^{3}}+\frac{6\dot{a}^{2}f^{\prime}%
(T)}{a^{2}N^{2}}-\frac{2kf^{\prime}(T)}{a^{2}}+\kappa\rho_{0}wa^{-3(w+1)}%
-\frac{1}{2}f(T)=0\\
&  T-\left(  6\left(  \frac{\dot{a}}{Na}\right)  ^{2}-\frac{6k}{a^{2}}\right)
=0.
\end{align}
The latter giving of course the expression for the torsion scalar which we see
in Eq. \eqref{Tscalar}.

Before proceeding, we need to mention that the use of Lagrangian \eqref{Lag} is not the unique starting point from which one can be led to the Hamiltonian description of this minisuperspace system. An alternative process, which is owed to Schutz \cite{Schutz}, would be to consider the degree of freedom of the perfect fluid, by introducing variables related to thermodynamic quantities. Later, the application of a canonical transformation at the Hamiltonian level, allows for the contribution of the fluid to be written as being linear in the momenta. This is an approach that has been widely used in the canonical quantization of minisuperspace systems \cite{Neto2,Neto1,Jala,Vakili2} and it leads to a Schr\"ondinger-like quantum equation for the corresponding system, where the fluid is used as an effective time parameter. This approach is more formal in the sense that it takes into account the perfect fluid degree of freedom, but the use of a classical canonical transformation can raise ambiguities in what regards the quantum equivalente between the two systems, before and after the transformation. To avoid this we choose here to use Lagrangian \eqref{Lag}; in any case, the resulting methodology in what regards the use of fractional derivatives, which we are going to see later, is not to be affected by this choice.

But, prior to discussing the Hamiltonian formulation of \eqref{Lag}, it is
useful to perform a comparison with the minisuperspace Lagrangians of $f(R)$
and $f(Q)$ cosmologies for a FLRW spacetime.

\subsection{Minisuperspace Lagrangians in the rest of the trinity}

A similar procedure can be followed in the case of $f(R)$ gravity with
\end{subequations}
\begin{equation}
\mathcal{A}_{grav}=-\frac{1}{2\kappa}\int\!\!\sqrt{-g}f(R)d^{4}x,
\end{equation}
and $g=\det g_{\mu\nu}$. Here, in order to reproduce the relevant
minisuperspace Lagrangian we have to also introduce, with a Lagrange multiplier,
the relation for the Ricci scalar as obtained by the - now nonzero- Riemann
curvature
\begin{equation}
R=6\left(  \frac{\ddot{a}}{aN^{2}}-\frac{\dot{a}\dot{N}}{aN^{3}}+\frac{\dot
{a}^{2}}{a^{2}N^{2}}+\frac{k}{a^{2}}\right)  .
\end{equation}
We avoid the cumbersome details since the procedure can be found easily in the
literature \cite{Vilenkin,Vakili,Huang,DimfR,AndfR}, and we just present the
end result for the minisuperspace Lagrangian, again in the case of a perfect
fluid matter source, with a linear barotropic equation of state, $p=w \rho$,
\begin{equation}%
\begin{split}
L_{f(R)}=  &  \frac{1}{2\kappa}\left[  \frac{6a\dot{a}^{2}f^{\prime}(R)}%
{N}+\frac{6a^{2}\dot{a}\dot{R}f^{\prime\prime}(R)}{N}-N\left(  a^{3}\left(
f(R)-Rf^{\prime}(R)\right)  +6kaf^{\prime}(R)\right)  \right] \\
&  +N\rho_{0}a^{-3w}.
\end{split}
\label{LagfR}%
\end{equation}
This Lagrangian is also valid in its context, it reproduces correctly the
result of the $f(R)$ gravity equations for a FLRW spacetime. Notice the
important difference in comparison to the Lagrangian \eqref{Lag}. The latter
has no velocity for the degree of freedom $T$ in contrast to \eqref{LagfR},
which possesses a velocity with respect to $R$. This substantial difference leads to
\eqref{Lag} requiring a modified treatment in the construction
of its Hamiltonian formulation than the usual treatment applied to Lagrangian
\eqref{LagfR} and to other minisuperspace Lagrangians.

In the case of $f(Q)$ gravity we start from
\begin{equation}
\mathcal{A}_{grav}=\frac{1}{2\kappa}\int\!\!\sqrt{-g}f(Q)d^{4}x.
\end{equation}
Let us require here, for reasons that we will explain shortly, that we treat
only the spatially flat case $k=0$. Then, if - among the three different
options for a connection - we choose the one which is not dynamical, then the
nonmetricity scalar can be written as
\begin{equation}
Q=6\left(  \frac{\dot{a}}{Na}\right)  ^{2}. \label{Qval}%
\end{equation}
For $f(Q)$ cosmology there is usually a variety of distinct connections which
are compatible with the equations of motion
\cite{Hohmann,Heis,Baha,fQus1,fQus2,ADe,fQus3,an2}. In the spatially flat case,
$k=0$, there exist three. However, only one of them is nondynamical, i.e. in the coordinate system where the line element is being given by \eqref{lineel}, the equation of motion for the connection is identically satisfied. This is the connection whose components vanish
(coincident gauge) in Cartesian coordinates and it is
only for this connection that you get relation \eqref{Qval}. For the other two
connections, in the coordinate system where the metric is diagonal, the
dynamical function related to the connection will also appear in the
expression for $Q$ \cite{Heis,fQus1}; meaning that you need to write a
Lagrangian that also produces Euler-Lagrange equations for the connection.
Anyway, let us consider the simplest case of the nondynamical connection. The
resulting minisuperspace Lagrangian is \cite{QuantfQus}
\begin{equation}
L_{f(Q)}=\frac{1}{2\kappa}\left[  \frac{6a\dot{a}^{2}f^{\prime}(Q)}{N}%
+Na^{3}\left(  f(Q)-Qf^{\prime}(Q)\right)  \right]  +N\rho_{0}a^{-3w}.
\label{LagfQ}%
\end{equation}
Here, there is also no velocity for the additional degree of freedom $Q$. This
Lagrangian is also successful in reproducing the result of the field equations
under the assumptions we have set $k=0$ and considering the nondynamical
connection. The Hamiltonian analysis and an attempt in the canonical
quantization of this Lagrangian, for various scenarios of matter content, has
been performed in \cite{QuantfQus}.

The similarity of Lagrangian \eqref{LagfQ} to that of the $f(T)$ theory, Eq.
\eqref{Lag}, for $k=0$ is obvious. The one becomes the other by interchanging
$T$ with $Q$. The same is true for the expressions of the basic scalars
\eqref{Tscalar} and \eqref{Qval}. At this point, one may be naively tempted to consider that,
in the $k\neq0$ case, the nonmetricity scalar is equal to the right-hand side
of \eqref{Tscalar} and write the minisuperspace Lagrangian of $f(Q)$ cosmology as \eqref{Lag} with
$T\rightarrow Q$; this is for example what has been done in \cite{Cap}. However, this
is incorrect since
\begin{equation}
Q\neq6\left(  \frac{\dot{a}}{Na}\right)  ^{2}-\frac{6k}{a^{2}},\quad
\text{when}\quad k\neq0.
\end{equation}
Unlike the $k=0$ case, in the coordinate systems where the FLRW metric is diagonal having an obvious
homogeneous and isotropic form, there is no nondynamical
connection when $k\neq0$. There exists one connection compatible with the symmetries of the spacetime; it is a nonzero, dynamical
connection, whose time dependence affects the expression of the nonmetricity
scalar $Q$.

It is a common problem in $f(Q)$ cosmology that one may be led to the error of thinking that the connection is
irrelevant, because its flatness implies the existence of a coordinate system where it becomes zero. However, as it has been also stressed in \cite{Zhao}, the adoption
of a specific type of spacetime metric, like the one in Eq. \eqref{lineel}, already
constitutes a partial gauge fixing. As a result, the coordinate system in
which we write the metric may not be compatible with having a zero connection. This is what happens in the $k\neq 0$ case.

So we see that, unlike the $f(T)$ case, in $f(Q)$ FLRW cosmology, the relevant
minisuperspace Lagrangian is not given by \eqref{Lag}, with a simple change
$T\rightarrow Q$, when $k\neq0$. It is only for $k=0$ where we can make this
identification. It can also be checked that, if one erroneously sets
$T\rightarrow Q$ in \eqref{Lag}, with $k\neq0$, and takes the relevant
Euler-Lagrange equations, then the latter are not even equivalent to the field
equations for $f(Q)$ theory reduced by the assumption of the case of a
spatially non-flat FLRW metric.

\section{Hamiltonian formalism for the $f(T)$ minisuperspace}

\label{sec4}

We start by giving a brief description of the usual Hamiltonian formalism for
minisuperspace Lagrangians in order to later see in what important way the
Lagrangian \eqref{Lag} differentiates from the usual prescription.
Minisuperspace Lagrangians are usually of the form
\begin{equation}
L=\frac{1}{2N}G_{ij}(q)\dot{q}^{i}\dot{q}^{j}-NV(q) \label{Lagminigen}%
\end{equation}
where $G_{ij}$ is the so called minisuperspace metric and $q^{i}$ are the rest
of the degrees of freedom, beside the lapse $N$ which we distinguish. In the
case of $f(R)$ gravity and Lagrangian \eqref{LagfR} we have $q^{i}=(a,R)$, in $f(Q)$ gravity
and \eqref{LagfQ} there is $q^{i}=(a,Q)$, and finally in the case that we want to
treat in this work, and Lagrangian \eqref{Lag}, we have $q^{i}=(a,T)$.

The fact that all Lagrangians have in common is that there is no velocity for
$N$. This already signifies that the Legendre transform is not invertible and
thus we need to invoke the Dirac-Bergmann algorithm in order to write the
Hamiltonian of the system, which ends up to be (given an invertible $G_{ij}$)
\begin{equation}
H_{T}=N\mathcal{H}+u_{N}p_{N}=N\left(  \frac{1}{2}G^{ij}(q)p_{i}%
p_{j}+V(q)\right)  +u_{N}p_{N}. \label{Hamminigen}%
\end{equation}
The momentum with respect to $N$, $p_{N}=\frac{\partial L}{\partial\dot{N}%
}\approx0$, is the primary constraint of the theory. We use the curly equal
\textquotedblleft$\approx$\textquotedblright\ in the case of constraints to
signify what Dirac mentioned as a weak equality. It basically means that we
are allowed to set the constraint equal to zero only when it appears outside
of Poisson brackets, i.e. $\{N,p_{N}^{2}\}=2\{N,p_{N}\}p_{N}=0$, but
$\{N,p_{N}\}=1$; even though $p_N=0$. The first expression is zero because there appears an overall
$p_{N}$ outside of a Poisson bracket. In the second example, it would be an error
to take $\{N,p_{N}\}=\{N,0\}=0$, i.e. set $p_N$ equal to zero inside a Poisson bracket. This is the meaning of denoting $p_N\approx 0$. Finally, an important remark is that, if a phase space quantity is weakly equal to zero, then it is (strongly) equal to a linear combination of the constraints of the system (at least in the case of systems with finite degrees of freedom like the ones we deal here).

Consistency requires that the constraints are to be preserved in time, at
least weakly, this means that we need to take
\begin{equation}
\dot{p}_{N}\approx0\Rightarrow\{p_{N},H_{T}\}\approx0\Rightarrow
\mathcal{H}\approx0.
\end{equation}
The
\begin{equation}
\mathcal{H}=\frac{1}{2}G^{ij}(q)p_{i}p_{j}+V(q)\approx0\label{genHamc}%
\end{equation}
is the secondary constraint of the theory and its own conservation in time is
identically satisfied. So, no additional constraints emerge. These two
constraints commute with each other, $\{p_{N},H\}=0$ and this classifies them
as being first class constraints. This is an important distinction, first
class constraints commute, at least weakly, with all of the constraints and
their existence is associated with some gauge freedom in the system. In
cosmology the existing manifest freedom is that of performing arbitrary
diffeomorphisms in time, which is what evidently remains at the
minisuperspace level from the full four-dimensional diffeomorphism
invariance\footnote{The automorphisms of the structure constants of the
spatial symmetries account for the freedom particular of mixed spatial-time
coordinate transformations \cite{ftn1}}. After having obtained the
Hamiltonian, the canonical quantization procedure by following Dirac's idea of
enforcing the constraints on the wave function leads to the equations
\begin{equation}
\hat{p}_{N}\psi=0,\quad\hat{\mathcal{H}}\psi=0.\label{quantcond}%
\end{equation}
The first equation is usually forgotten most of the time, together with $N$
all-together in \eqref{Lagminigen} and $p_{N}$ in \eqref{Hamminigen}. But it
is the existence of the $p_{N}\approx0$, which leads us to $\mathcal{H}\approx 0$ and later to
$\hat{\mathcal{H}}\psi=0$, which is the Wheeler-DeWitt equation. The first of
\eqref{quantcond}, $\hat{p}_{N}\psi=0\Rightarrow-i \hbar \frac{\partial\psi}{\partial
N}=0$, just tells us that the wave function $\psi$ does not depend on the
lapse $N$.

The above procedure is perfectly fine for Lagrangians like \eqref{LagfR}.
However it is not sufficient for the cases of $f(Q)$ and of $f(T)$
cosmologies. The reason lies in the fact that for \eqref{LagfQ} and
\eqref{Lag}, the corresponding $G_{ij}(q)$ matrix is not invertible. We
already noticed the reason for this; it is the lack of velocity terms in the
Lagrangian for $Q$ and $T$ respectively.

Let us concentrate in Lagrangian \eqref{Lag}, whose quantization we want to
study. As we mentioned, in \eqref{Lag}, we are not only missing a velocity for
$N$, we are also missing a velocity for $T$, i.e. there is no $\dot{T}$ term.
So, according to the Dirac-Bergmann algorithm, instead of one, we start off
with two primary constraints
\begin{equation}
p_{N}=\frac{\partial L}{\partial\dot{N}}\approx0,\quad p_{T}=\frac{\partial
L}{\partial\dot{T}}\approx0.
\end{equation}
We may write the canonical part $NH$ of the Hamiltonian with what remains of
the degrees of freedom - so we have
\begin{equation}
\dot{a}p_{a}-L=N\mathcal{H}=N\left[  \frac{\kappa p_{a}^{2}}{12af^{\prime}%
(T)}-\frac{a^{3}}{2\kappa}\left(  f(T)-Tf^{\prime}(T)\right)  +\frac
{3k}{\kappa}af^{\prime}(T)-\rho_{0}a^{-3w}\right]  .
\end{equation}
The total Hamiltonian is obtained by adding to it the primary constraints with
multipliers
\begin{equation}
H_{T}=N\mathcal{H}+u_{N}p_{N}+u_{T}p_{T}. \label{Htot}%
\end{equation}
As explained before, consistency requires now that both $p_{N}$ and $p_{T}$
are preserved in time, thus we need to require
\begin{equation}
\{p_{N},H_{T}\}\approx0\quad\text{and}\quad\{p_{T},H_{T}\}\approx0.
\end{equation}
The first relation leads to the expected Hamiltonian constraint, because again
$\{p_{N},H_{T}\}=-\mathcal{H}$, so we need to impose
\begin{equation}
\mathcal{H}=\frac{\kappa p_{a}^{2}}{12af^{\prime}(T)}-\frac{a^{3}}{2\kappa
}\left(  f(T)-Tf^{\prime}(T)\right)  +\frac{3k}{\kappa}af^{\prime}(T)-\rho
_{0}a^{-3w}\approx0. \label{Hamcon}%
\end{equation}
The second consistency condition is
\begin{equation}
\{p_{T},H_{T}\}=p_{a}^{2}-\frac{6}{\kappa^{2}}a^{2}f^{\prime}(T)^{2}\left(
a^{2}T+6k\right)  \approx0.
\end{equation}
This is nothing more than relation \eqref{Tscalar} expressed in the phase
space formalism and it forms an additional secondary constraint together with
the $\mathcal{H}\approx0$ of \eqref{Hamcon}. We symbolize this new constraint with
$\varphi$,
\begin{equation}
\varphi=p_{a}^{2}-\frac{6}{\kappa^{2}}a^{2}f^{\prime}(T)^{2}\left(
a^{2}T+6k\right)  \approx0. \label{secondary2}%
\end{equation}
The further checking of the consistency conditions $\{\mathcal{H},H_{T}\}\approx0$ and
$\{\varphi,H_{T}\}\approx0$ leads to no additional constraints. The first is
proportional to $\varphi\approx0$ so it is weakly zero
\begin{equation}
\{\mathcal{H},H_{T}\}=\frac{\kappa Nf^{\prime\prime}(T)}{12af^{\prime}(T)^{2}%
}\varphi\approx0,
\end{equation}
while $\{\varphi,H_{T}\}\approx0$ leads to the derivation of the multiplier
$u_{T}$
\begin{equation}
\{\varphi,H_{T}\}\approx0\Rightarrow u_{T}\approx\frac{\kappa Np_{a}\left(
f(T)-2Tf^{\prime}(T)-4kf^{\prime}(T)a^{-2}-2\kappa\rho_{0}wa^{-3(w+1)}\right)
}{2f^{\prime}(T)\left(  a^{2}\left(  2Tf^{\prime\prime}(T)+f^{\prime
}(T)\right)  +12kf^{\prime\prime}(T)\right)  }, \label{velT}%
\end{equation}
if we assume of course that the denominator is different from zero, i.e.
\begin{equation}
a^{2}\left(  2Tf^{\prime\prime}(T)+f^{\prime}(T)\right)  +12kf^{\prime\prime
}(T)\neq0. \label{denominator}%
\end{equation}
In the expression for $u_{T}$ we have also used relation \eqref{secondary2} to
eliminate $p_{a}^{2}$ and simplify the end result.

The branch that opens when the expression of the left hand side of
\eqref{denominator} is zero is quite interesting and significantly
differentiates the procedure to be followed. For this reason, we will study
separately this particular case, where $u_{T}$ is not fixed, but first let us
treat the generic case.

\subsection{Generic case} \label{secfrac}

If we assume that \eqref{denominator} holds, we have a four-dimensional set of constraints, spanned by
\[
\{p_{N},p_{T},\mathcal{H},\varphi\}.
\]
The fact that one of the multipliers appearing in \eqref{Htot} is fixed,
signifies the existence of second class constraints, that is, constraints
whose Poisson brackets are not zero even in the weak sense. This implies the
existence of ``fake'' or non-physical degrees of freedom in the system. The second class
constraints in the above set are the $p_{T}$ and $\varphi$, whose Poisson bracket
yields
\begin{equation}
\{p_{T},\varphi\}=\frac{6a^{2}}{\kappa^{2}}f^{\prime}(T)\left(  a^{2}\left(
2Tf^{\prime\prime}(T)+f^{\prime}(T)\right)  +12kf^{\prime\prime}(T)\right) \neq 0 .
\label{secclpoi}%
\end{equation}
The above expression being nonzero due to \eqref{denominator}, which we assume that holds in the generic case.  Exactly because $p_{T}$ and $\varphi$ are second class, we cannot proceed with
the quantization in the same manner we do for first class constraints. For
example, the fact that $p_{T}\approx0$ cannot imply now at the quantum level
$\hat{p}_{T}\phi=0$, as it happens for the $p_{N}\approx0$ case. The reason is
that this would lead to an inconsistency due to \eqref{secclpoi} being
nonzero. We would have at the quantum level $\hat{p}_{T}\psi=0$ and
$\hat{\varphi}\psi=0$ with $[\hat{p}_{T},\hat{\varphi}]\psi$ being nonzero by construction,
which is impossible. There are two main strategies regarding this predicament.
The one requires the enhancement of the phase space with the addition of extra
degrees of freedom in such a way so that the second class constraints become
first class. In this bigger phase space one can proceed with the usual
quantization of first class constraints. The construction however of the
necessary Hamiltonian, with the extra degrees of freedom, that will reproduce
the same dynamics, is far from trivial.

The second approach, which was what was initially proposed by Dirac, is the
elimination of the fake degrees of freedom represented by the second class
constraints. This is done with the introduction of the Dirac bracket in place
of the Poisson. For two phase space functions $A$, $B$, the former is defined
as
\begin{equation}
\{A,B\}_{D}=\{A,B\}-\{A,X_{i}\}\Delta_{ij}^{-1}\{X_{j},B\}, \label{Dbra}%
\end{equation}
where $X$s are the elements of the set of second class constraints and
$\Delta^{-1}$ is the inverse matrix of the one constructed by the Poisson
brackets of the second class constraints. In our case the set of second class
constraints is two dimensional (so $i,j=1,2$) and made up by $X_{1}=p_{T}$,
$X_{2}=\varphi$. The antisymmetric $2\times2$ matrix $\Delta$, whose inverse
appears in \eqref{Dbra}, has components $\Delta_{ij}=\{X_{i},X_{j}\}$. The initial idea behind the introduction of Dirac brackets was to impose all second class constraints strongly equal to zero by making them commute with any phase space function. As a trivial byproduct of this, all constraints, irrespectively of being first or second class, commute with each other through the use of Dirac bracket.

At this point we can enforce $p_{T}=0$ and $\varphi=0$, as strong equations. That
is, substitute inside the Hamiltonian $p_{T}=0$ and $p_{a}$ from
\eqref{secondary2}, thus arriving at the reduced Hamiltonian
\begin{equation}
H_{T}^{red}=N\mathcal{H}_{red}+u_{N}p_{N} \label{Hamred}%
\end{equation}
where the reduced, stripped from fake degrees of freedom, Hamiltonian
constraint is
\begin{equation}
\mathcal{H}_{red}=\frac{1}{2\kappa}\left[  a^{3}\left(  2Tf^{\prime
}(T)-f(T)\right)  +12kaf^{\prime}(T)\right]  -\rho_{0}a^{-3w}\approx0.
\end{equation}
The reduced system has only the two remaining first class constraints
$p_{N}\approx0$ and $H_{red}\approx0$ and can be quantized in the usual
manner, with the exception now that the canonical quantization scheme is to be
constructed based on the Dirac brackets and not the Poisson. That is, we need
to find functions on the reduced phase space $q(a,T)$ and $p(a,T)$ with the
property
\begin{equation}
\{q(a,T),p(a,T)\}_{D}=1. \label{Diraccanonical}%
\end{equation}
These are going to be our canonical variables to which we will need to assign
the basic quantum operators of position $\hat{q}\psi=q\psi$ and momentum
$\hat{p}\psi=-i \frac{d}{dq}\psi$ (from now on we assume units $\hbar=1$).

For $k=0$, a very convenient choice of functions satisfying
\eqref{Diraccanonical} is
\begin{equation}
q(a,T)=\pm\left(  2Tf^{\prime}(T)-f(T)\right)  \quad\text{and}\quad
p(a,T)=\frac{a^{3}}{\kappa\sqrt{6T}}. \label{canvarkzero}%
\end{equation}
In particular we have
\begin{equation}
\{q(a,T),p(a,T)\}_{D}=\mp\frac{\kappa p_{a}}{\sqrt{6T}a^{2}f^{\prime}(T)}=1,
\end{equation}
with the last equation being true by virtue of the constraint
\eqref{secondary2}, which we substitute as $p_{a}=\mp\sqrt{6T}a^{2}f^{\prime
}(T)/\kappa$. Note that the arbitrariness in the sign of $p_{a}=\frac{\partial
L}{\partial\dot{a}}=6af^{\prime}(T)\dot{a}/(\kappa N)$ is related with the
arbitrariness in the sign of $N$ (only $N^{2}$ appears in the metric). This
carries over an arbitrariness in the sign that we may use to define $q$ in
\eqref{canvarkzero}. As we are going to see this does not really affect the
end result, but it may be useful in allowing us to expand the domain of definition
of the $q$ variable depending on the $f(T)$ theory that we adopt.

In the new variables the reduced Hamiltonian constraint reads
\begin{equation}
\mathcal{H}_{red}=\pm\sqrt{\frac{3}{2}}q\sqrt{T(q)}p-6^{-\frac{w}{2}}\rho
_{0}\kappa^{-w}T(q)^{-\frac{w}{2}}p^{-w}\approx0, \label{conred}%
\end{equation}
where $T(q)$ is the inverse of the first of \eqref{canvarkzero}, which is to
be found for some given $f(T)$ theory. We can exploit the
time-reparmaterization invariance of the system in order to scale the lapse
function as
\begin{equation}
N\mapsto\bar{N}=\sqrt{\frac{3}{2}}T^{-\frac{w}{2}}p^{-w}N, \label{rescaleN}%
\end{equation}
which allows us to write the reduced Hamiltonian \eqref{Hamred} as
\begin{equation}
H_{T}^{red}=\bar{N}\bar{\mathcal{H}}_{red}+u_{\bar{N}}p_{\bar{N}},
\label{Hamredscaled}%
\end{equation}
where we introduced the re-scaled Hamiltonian constraint
\begin{equation}
\bar{\mathcal{H}}_{red}=\pm qT(q)^{\frac{w+1}{2}}p^{1+w}-\frac{2^{\frac
{1-w}{2}}\rho_{0}}{3^{\frac{1+w}{2}}\kappa^{w}}\approx0. \label{conredscaled}%
\end{equation}
Having all the dynamical dependence incorporated in one term helps
significantly in the construction of an appropriate quantum operator for the
$\bar{\mathcal{H}}_{red}$ and allows us to interpret the quantum version of
\eqref{conredscaled} as an eigenvalue equation. Note that the scaling we did
in \eqref{rescaleN} is equivalent to a change in the lapse in the metric of
the form $N=\bar{N}\frac{a^{3w}}{6^{w/2}\kappa^{w}}$ due to $p$ being given by \eqref{canvarkzero}.

From what we observe in \eqref{conredscaled}, the momentum is raised in a
power which is not necessarily a natural number. Such Hamiltonians are the
object of study in the theory of fractional quantum mechanics \cite{Laskin1,Laskin2,Laskin3}. Several
different implementations of fractional derivatives exist in the literature \cite{frd1,frd2}. In
\cite{QuantfQus} we made use of the Katugampola fractional derivative, which has been introduced
in \cite{Katuga}. This derivative has the advantage of producing a differential
equation for the quantum version of \eqref{conredscaled}, instead of an
integro-differential equation as happens in the case of other fractional derivatives.
This significantly simplifies the derivation of the relative quantum solutions.
The Katugampola fractional derivative with index $\alpha$ is defined as
\begin{equation}
D^{\alpha}(\Psi(q))=\underset{\epsilon\rightarrow0}{\lim}\frac{\Psi
(qe^{\epsilon q^{-\alpha}})-\Psi(q)}{\epsilon}, \label{Katu0}%
\end{equation}
when $0<\alpha\leq1$ and $q>0$. Here, the arbitrariness of the sign in the
definition of $q$ in \eqref{canvarkzero} can come into play, since for
different regions in the domain of $T$ we may choose that definition of $q$ so that the latter remains positive ($q>0$)
in the relative domain. Remember that we have the ability to define $q$ from
\eqref{canvarkzero} with different overall signs with respect to the
expression $2Tf^{\prime}(T)-f(T)$, depending on whether the latter is positive
or negative.

The action of \eqref{Katu0} upon a given function gives
\begin{equation}
D^{\alpha}(\Psi(q))=q^{1-\alpha}\frac{d\Psi}{dq}. \label{Katu1}%
\end{equation}
Although, initially the derivative is introduced for $0<\alpha\leq1$, the
definition can be modified to be extended for other values of $\alpha$. For
example, for $1<\alpha\leq2$ one obtains the action
\begin{equation}
D^{\alpha}(\Psi(q))=q^{2-\alpha}\frac{d^{2}\Psi}{dq^{2}}. \label{Katu2}%
\end{equation}
These are the two operators that are of interest to us.

We have a Hamiltonian constraint given by Eq. \eqref{conredscaled}, which is of the general form
\begin{equation} \label{conredscaledsimp}
\bar{\mathcal{H}}_{red}=\pm A(q)p^{\alpha}-\Sigma^{2}\approx0,
\end{equation}
where
\begin{equation}
A(q)=qT(q)^{\frac{w+1}{2}},\quad\alpha=1+w\quad\text{and}\quad\Sigma^{2}%
=\frac{2^{\frac{1-w}{2}}\rho_{0}}{3^{\frac{1+w}{2}}\kappa^{w}}. \label{AaS}%
\end{equation}

When the power of the momentum $\alpha=1+w$ is in the region $0<\alpha\leq1$,
i.e. $-1<w\leq0$, we assign to $\bar{\mathcal{H}}_{red}$ the operator
\begin{equation}
\widehat{\mathcal{H}}=\mp\frac{\mathrm{i}}{2\mu(q)}\left[  \mu
(q)A(q)q^{1-\alpha}\frac{d}{dq}+\frac{d}{dq}\left(  \mu(q)A(q)q^{1-\alpha
}\;\right)  \right]  -\Sigma^{2}. \label{operator1}%
\end{equation}
The $\widehat{\mathcal{H}}$ of \eqref{operator1} has been constructed so that
at the limit $\alpha=1$ it gives the most general, linear, first order
differential operator which is Hermitian under a measure $\mu(q)$. The
equivalent of the Wheeler-DeWitt equation in this case is given by taking
$\widehat{\mathcal{H}}\Psi=0$ which results to the eigenvalue equation
\begin{equation}
\mp\frac{\mathrm{i}}{2\mu(q)}\left[  \mu(q)A(q)q^{1-\alpha}\frac{d\Psi}%
{dq}+\frac{d}{dq}\left(  \mu(q)A(q)q^{1-\alpha}\Psi\right)  \right]
=\Sigma^{2}\Psi . \label{eigen1}%
\end{equation}
The latter has the general solution
\begin{equation}
\Psi=C\frac{q^{\frac{\alpha-1}{2}}}{\mu(q)^{\frac{1}{2}}A(q)^{\frac{1}{2}}%
}\exp\left[  \pm\mathrm{i}\,\Sigma^{2}\int\frac{q^{\alpha-1}}{A(q)}dq\right]
, \label{Psi1}%
\end{equation}
where $C$ denotes the constant of integration and the $A(q)$, $\alpha$ and
$\Sigma$ are given by \eqref{AaS}. We can write the corresponding probability
amplitude to be
\begin{equation}
P=\mu(q)\Psi^{\ast}\Psi=CC^{\ast}\frac{q^{\frac{\alpha-1}{2}}}{A(q)^{\frac
{1}{2}}}\left(  \frac{q^{\frac{\alpha-1}{2}}}{A(q)^{\frac{1}{2}}}\right)
^{\ast}=\frac{|C|^{2}}{T^{\frac{1+w}{2}}|2Tf^{\prime}(T)-f(T)|^{1-w}},
\label{proampl0}%
\end{equation}
where in the last equality we used the definition of $q$, as a function of
$T$, from \eqref{canvarkzero}. Note that choosing a particular expression for
the measure function in \eqref{Psi1} is irrelevant, since it is eliminated from the final expression
of the probability amplitude. Of course, for a well-defined probabilistic
interpretation, the integral of $P$, for all $q$ (or similarly $T$), should be
finite. However, as usually happens in cosmology, this is not in general the
case. Nevertheless, we are able to provide a normalization up to a delta
function if we write, for two different values $\Sigma$ and $\Sigma^{\prime}$,
the inner product
\[
\int\!\!\mu(q)\Psi_{\Sigma}(q)\Psi_{\Sigma^{\prime}}^{\ast}(q)dq=|C|^{2}%
\int\!\!\frac{q^{\alpha-1}}{A(q)}e^{\left[  i\int\!\!\frac{q^{\alpha-1}}%
{A(q)}dq\left(  \Sigma^{2}-\Sigma^{\prime2}\right)  \right]  }dq=|C|^{2}%
\int\!\!e^{\mathrm{i}(\Sigma^{2}-\Sigma^{\prime2})u}du,
\]
where we performed the change of variable $u=\int\!\!\frac{q^{\alpha-1}}%
{A(q)}dq$. If the variable $u$ in the last integral runs in the whole real line, then
he above expression becomes equal to $2\pi|C|^{2}\delta(\Sigma^{2}%
-\Sigma^{\prime2})$ providing a normalization value for the $C$ constant,
$C=\frac{1}{\sqrt{2\pi}}$. Of course, the domain of definition of $q$ and
consequently of $u$ depends on the assumed $f(T)$ theory that we consider.

From what we observe in Eq. \eqref{proampl0}, we expect the probability amplitude
to favour values of $T$ that lie on the roots of $T^{\frac{1+w}{2}%
}(2Tf^{\prime}(T)-f(T))^{1-w}=0$. Take for example theories that imply a
power-law modification added to the original GR dynamics, that is
$f(T)=T+\beta T^{\mu}$. Then, we see that the probability amplitude goes to
infinity if
\begin{equation}
T^{\frac{1+w}{2}}(2Tf^{\prime}(T)-f(T))^{1-w}=T^{\frac{3-w}{2}}\left(
\beta(2\mu-1)T^{\mu-1}+1\right)  ^{1-w}=0.
\end{equation}
Assuming that $T\geq0$ and given that the amplitude corresponds to the case
$-1<w\leq0$, the above equation is satisfied at $T=0$, and if $\beta<0$ and
$\mu>0$, also at $T=\left[  \beta(1-2\mu)\right]  ^{\frac{1}{1-\mu}}$. We are going to see some examples of this in the subsequent analysis.

Now, if the power of the momentum is $1<\alpha\leq2$ in \eqref{conredscaledsimp}, that is $0<w\leq1$ for \eqref{conredscaled}, we
use as the Hamiltonian constraint operator
\begin{equation}
\widehat{\mathcal{H}}=\mp\frac{1}{\mu(q)}\frac{d}{dq}\left[  \mu
(q)A(q)q^{2-\alpha}\frac{d}{dq}\;\right]  -\Sigma^{2}, \label{operator2}%
\end{equation}
in which the measure function is $\mu=|q^{\frac{\alpha}{2}-1}A(q)^{-\frac
{1}{2}}|$. The differential part of the above expression, for $\alpha=2$,
becomes the one dimensional Laplacian, which we would normally use in the case of a quadratic
Hamiltonian. The general solution to the eigenvalue equation
\begin{equation}
\mp\frac{1}{\mu(q)}\frac{d}{dq}\left[  \mu(q)A(q)q^{2-\alpha}\frac{d}{dq}%
\Psi\right]  =\Sigma^{2}\Psi, \label{eigen2}%
\end{equation}
is
\begin{equation}
\Psi=C_{1}\exp\left[  (\mp1)^{\frac{1}{2}}\Sigma\int\frac{q^{\frac{\alpha}%
{2}-1}}{A(q)^{\frac{1}{2}}}dq\right]  +C_{2}\exp\left[  -(\mp1)^{\frac{1}{2}%
}\Sigma\int\frac{q^{\frac{\alpha}{2}-1}}{A(q)^{\frac{1}{2}}}dq\right]  ,
\label{Psi2}%
\end{equation}
which, in the case of the upper sign $(-1)^{\frac{1}{2}}=i$, can be seen as a
linear combination of \textquotedblleft ingoing\textquotedblright\ and
\textquotedblleft outgoing\textquotedblright\ waves. If we choose either one
of the two waves we can impose a normalizability up to a delta function. For
example, take for two different values $\Sigma$ and $\Sigma^{\prime}$ the
integral
\begin{equation}
\int\!\!\mu(q)\Psi_{\Sigma}(q)\Psi_{\Sigma^{\prime}}^{*}(q)dq=|C_{1}|^{2}%
\int\!\!e^{\mathrm{i}(\Sigma-\Sigma^{\prime})x}dx,
\end{equation}
where we have set $x=\int\!\!\frac{q^{\frac{\alpha}{2}-1}}{A(q)^{\frac{1}{2}}%
}dq$. The above integral results in $2\pi|C_{1}|^{2}\delta(\Sigma
-\Sigma^{\prime})$ if we consider now the $x$ to run in the whole set of the real numbers.

Similarly to \eqref{proampl0}, the probability amplitude, for a single state
given by the first branch of \eqref{Psi2}, becomes
\begin{equation}
P=\mu(q)\Psi^{\ast}\Psi=C_{1}C_{1}^{\ast}\frac{q^{\frac{\alpha}{2}-1}%
}{A(q)^{\frac{1}{2}}}=\frac{|C_{1}|^{2}}{T^{\frac{1+w}{2}}\left(  2Tf^{\prime
}(T)-f(T)\right)  ^{\frac{2-w}{2}}}, \label{proampl0b}%
\end{equation}
which again we may claim that it can assume its maximum (diverging) values at the points where
the denominator goes to zero. This holds for the upper sign of \eqref{eigen2}
and \eqref{Psi2}, which corresponds to having $q=2Tf^{\prime}(T)-f(T)>0$. When
$2Tf^{\prime}(T)-f(T)<0$ however, we use the opposite definition for $q$ from
\eqref{canvarkzero} with $q=-\left(  2Tf^{\prime}(T)-f(T)\right)  >0$, which
leads to real exponents in \eqref{Psi2}. The probability amplitude in this
case can be taken to be
\begin{equation}
P=\mu(q)\Psi^{\ast}\Psi=|C_{2}|^{2}\frac{q^{\frac{\alpha}{2}-1}}%
{A(q)^{\frac{1}{2}}}\exp\left[  -2\Sigma\int\frac{q^{\frac{\alpha}{2}-1}%
}{A(q)^{\frac{1}{2}}}dq\right]  , \label{proampl0c}%
\end{equation}
where we have chosen the branch that (in the following examples) decays as
$T\rightarrow\infty$.

\begin{figure}[ptb]
{\normalsize
\includegraphics[width=1\textwidth]{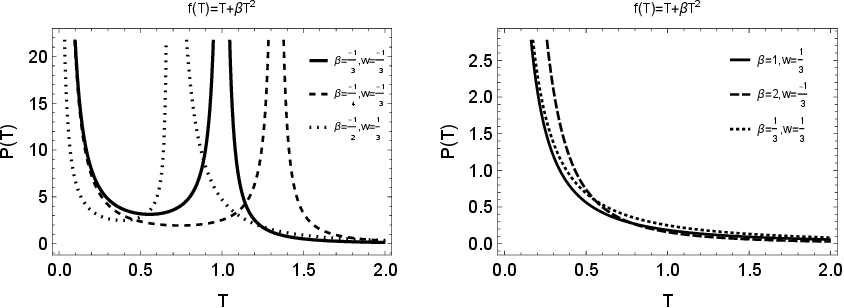} }\caption{The probability
amplitude $P$ for different values of $\beta$ and $w$ in the $f(T)=T+\beta
T^{2}$ theory. One can distinguish between two ($T=0$, $T=T_{0}\neq0$) and a
single ($T=0$) most probable values for the torsion scalar $T$. Expressions
\eqref{proampl0}, \eqref{proampl0b} and \eqref{proampl0c} have been used for
$P$, with $C=C_{1}=C_{2}=\frac{1}{\sqrt{2\pi}}$ respectively and, in the case
of \eqref{proampl0c}, we have additionally set, for reasons of facilitating
the depiction of the graphs, $\Sigma=1/6$.}%
\label{fig0a}%
\end{figure}

In Fig. \ref{fig0a} we present the graphs of the probability amplitudes, in
various cases, for theories of the form $f(T)=T+\beta T^{2}$. In the first
part of Fig. \ref{fig0a}, the fact that we take $\beta<0$ leads the common
denominator in \eqref{proampl0}, \eqref{proampl0b} and \eqref{proampl0c} to
become zero at two points: $T=0$ and $T=-\frac{1}{3\beta}$; we see
the probability amplitude diverging at the respective points. Notice that for
the cases $w=-1/3$, \eqref{proampl0} has been used since it belongs to the
case $0<\alpha\leq1$, while for $w=1/3$ ($1<\alpha\leq2$) we used
\eqref{proampl0b} when $2Tf^{\prime}(T)-f(T)>0$ (that is $0<T<-\frac{1}%
{3\beta}$) and \eqref{proampl0c} when $2Tf^{\prime}(T)-f(T)<0$, i.e.
$T>-\frac{1}{3\beta}$. From the definition of the fractional derivative we use we need to have $q>0$. One needs to be careful in the calculation of the various probability amplitudes regarding this point. The $q$ introduced in Eq. \eqref{canvarkzero}, has in its definition an arbitrariness in the sign which does not affect the functional form of the wave function. We make use of this to split regions of $T$, where we assign the appropriate expression for $q$, from \eqref{canvarkzero}, which remains positive in that region.

In the second set of graphs appearing in Fig. \ref{fig0a} we depict cases with
positive $\beta$, here the highest probability amplitude is encountered in the
region $T\rightarrow0$. Unlike the $\beta<0$ case of the first part, here
there is only one dominating value, implying that the torsion scalar is expected to be
almost zero.

It is interesting to note that the classical equations \eqref{euleqs}, for
$k=0$ and for $f(T)=T+\beta T^{2}$, are solved by the combination
\begin{equation}
N^{2}=\frac{6\dot{a}^{2}}{a^{2}T},\quad T=\frac{-1\pm\sqrt{1+24\beta\kappa
\rho_{0}a^{-3(1+w)}}}{6\beta}. \label{solsq}%
\end{equation}
The latter implies that, for $\beta<0$, and as long as $w>-1$, the torsion
scalar $T$ is bounded by $0\leq T\leq-\frac{1}{3\beta}$. Both values are
reached asymptotically as $a\rightarrow+\infty$; the first for the upper sign
appearing in the expression for $T$ in \eqref{solsq} and the other for the
lower. Thus, from what we may observe in Fig. \ref{fig0a}, there is a quantum
region assigning nonzero amplitudes at classically restricted values of $T$; the region $T>-\frac{1}{3\beta}$ in the first set of plots.

\begin{figure}[ptb]
{\normalsize
\includegraphics[width=1\textwidth]{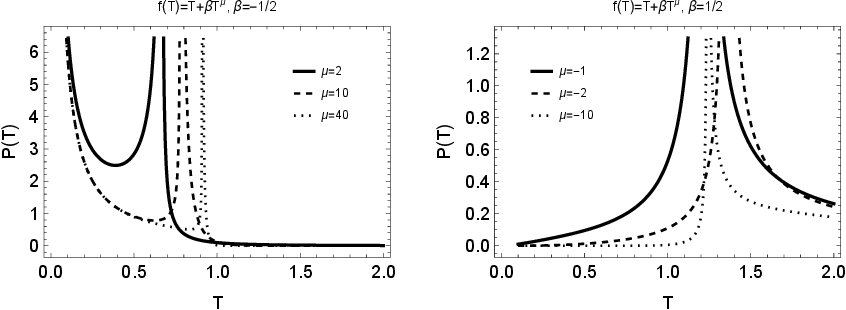} }\caption{Probability
amplitude $P$ in theories $f(T)=T+\beta T^{\mu}$ for different values of $\mu
$. The first set of plots corresponds to $\mu>0$, leading to two most probable
values, while in the second case we can see theories with $\mu<0$ leading to a
single dominant non-zero value for $T$. Once more, expressions
\eqref{proampl0}, \eqref{proampl0b} and \eqref{proampl0c} have been used
appropriately, with $C=C_{1}=C_{2}=\frac{1}{\sqrt{2\pi}}$ and $\Sigma=1/6$.}%
\label{fig0b}%
\end{figure}

In Fig. \ref{fig0b} we generalize the previous analysis in the case of
$f(T)=T+\beta T^{\mu}$ theories. Here we see how the different values of $\mu$
affect the probability amplitude. The first set of graphs is similar to what
we have seen before in Fig. \ref{fig0a}; theories which lead to two distinct
most probable values for the torsion scalar $T$. In the second plot we see
theories with negative $\mu$ in which $T=0$ no longer maximizes the
probability amplitude. There is a finite nonzero value of $T$ with that property. Similarly to Fig. \ref{fig0a}, we have used \eqref{proampl0},
\eqref{proampl0b} or \eqref{proampl0c} for the relevant regions of $T$ and the
values of the parameters.

In both Figures \ref{fig0a} and \ref{fig0b} we recognize as a relative good
sign the fact that in all cases $P(T)\rightarrow0$ as $T\rightarrow+\infty$ (of course, in the $0<w \leq 1$ case, we chose the branch of the wave function which would guarantee this property). This in a sense assigns a zero probability amplitude to values that would
correspond to a divergence of the geometric scalar of the theory, the torsion
scalar $T$. Avoiding thus, the \textquotedblleft equivalent\textquotedblright of a classical curvature singularity which, is encountered in metric theories of gravity.

The analysis we performed up to now was dedicated to the generic case, where
\eqref{denominator} holds, but strictly for $k=0$. The generic case of $k\neq 0$ is plagued by an important difficulty. Although it is quite easy to find a pair of canonical variables with respect to Dirac brackets satisfying Eq. \eqref{Diraccanonical}, it is highly nontrivial to produce a Hamiltonian which can be straightforwardly assigned to a quantum operator. One usually obtains complicate functions $\mathcal{H}(q,p)$ for which we do not currently have a clear prescription to propose on how to proceed.

In the next section we treat the particular case, where \eqref{denominator} does not hold and the relevant expression is
zero. As we are going to see, this case makes sense (for modified dynamics from GR) only if $k\neq0$. In order to avoid the general complication of the $k\neq0$ case, we shall follow a different procedure, exploiting the fact that classically you can easily obtain a relation between $T$ and $a$.

\subsection{The particular case}

In this section we study what happens when
\begin{equation}
a^{2}\left(  2Tf^{\prime\prime}(T)+f^{\prime}(T)\right)  +12kf^{\prime\prime
}(T)=0. \label{denominatorzero}%
\end{equation}
First, let us notice that this case if of interest only when $k\neq0$. If
$k=0$, the reduced expression
\begin{equation}
\left(  2Tf^{\prime\prime}(T)+f^{\prime}(T)\right)  =0
\label{denominatorzerokzero}%
\end{equation}
can be satisfied either if $f(T)\sim\sqrt{T}$, which leads to a Lagrangian
which is a total derivative, or if $T$ is a constant whose value is such that
satisfies algebraically equation \eqref{denominatorzerokzero}. In this latter
case, the resulting spacetime is the well-known de Sitter solution of General
relativity, where now the cosmological constant is given in terms of $T$. As a
result, we see that interesting novel solutions may emerge only in the context
of $k\neq0$, and this is what we will study here.

The treatment that we will apply is different than before. This has to do with the enhanced complication that the factor $k\neq 0$ brings into play. Even though we can find some canonical conjugate variables with respect to Dirac brackets, the Hamiltonian becomes so involved in these variables, that we do not have a clear recipe on how to map them to quantum operators. We shall thus exploit here the fact that we can relate easily $a$ to $T$ at the classical level.

If we use \eqref{denominatorzero}, together with the known expression for the
torsion scalar \eqref{Tscalar}, we can derive from the former an expression
for the lapse
\begin{equation}
N^{2}=-\frac{12\dot{a}^{2}f^{\prime\prime}(T)}{a^{2}f^{\prime}(T)}.
\label{partlapse}%
\end{equation}
By utilizing \eqref{denominatorzero}, its time derivative, and the lapse given by \eqref{partlapse}, in the
Euler-Lagrange equations \eqref{euleqs}, we arrive at the supplementary
conditions
\begin{subequations}
\label{supplcondf}%
\begin{align}
f^{\prime}(T)  &  =\frac{a^{2}f(T)+2\kappa\rho_{0}a^{-3w-1}}{2a^{2}%
T+12k}\label{supplcondf2}\\
f(T)  &  =\frac{\kappa\rho_{0}a^{-3(w+1)}}{2k}\left[ (w+1)a^{2}T+2k(3w+1)\right] %
\end{align}
We now take the time derivative of the last relation, divide with $\dot{T}$,
and set it equal to the first relation. Thus, we obtain a differential equation
connecting $T$ with $a$
\end{subequations}
\begin{equation}
2(3w+1)\dot{a}\left(  a^{2}T+6k\right)  -a^{3}\dot{T}=0.
\end{equation}
It's solution is
\begin{equation}
T(a)=%
\begin{cases}
-\frac{6(3kw+k)}{(3w+2)a^{2}}+T_{0}a^{2(3w+1)}, & \mbox{when }w\neq-\frac
{2}{3}\\
\frac{T_{0}}{a^{2}}-12k\frac{\ln a}{a^{2}}, & \mbox{when }w=-\frac{2}{3}%
\end{cases}
, \label{parttorsc}%
\end{equation}
where $T_{0}$ is a constant of integration. Up to now we have guaranteed the
consistency of \eqref{supplcondf}, we need also to check the consistency with
\eqref{denominatorzero} and of course with the Euler-Lagrange equations. This
can be done easily if we use \eqref{parttorsc} to write the $f(T)$ as a
function of the scale factor $f(a)$. From \eqref{supplcondf2}, we obtain
\begin{equation}
f(a)=%
\begin{cases}
\frac{T_{0}\kappa\rho_{0}(w+1)a^{3w+1}}{2k}-\frac{\kappa\rho_{0}%
(3w+1)a^{-3(w+1)}}{3w+2}, & \mbox{when }w\neq-\frac{2}{3}\\
\frac{\kappa\rho_{0}}{a}\left(  \frac{T_{0}}{6k}-1-2\ln a\right)  , &
\mbox{when }w=-\frac{2}{3}.
\end{cases}
. \label{partfa}%
\end{equation}
By making the change $f(T)\rightarrow f(a)$ in the Euler-Lagrange equations we
can easily see that the latter are satisfied with the substitution of
\eqref{partlapse}, \eqref{parttorsc} and \eqref{partfa}. From the first, we obtain the
expression for the lapse function
\begin{equation}
N(t)=%
\begin{cases}
\left[  \frac{6(3w+2)}{T_{0}(3w+2)a^{2(3w+2)}+6k}\right]  ^{\frac{1}{2}}%
\dot{a}, & \mbox{when }w\neq-\frac{2}{3}\\
\left(  \frac{6}{T_{0}+6k-12k\ln a}\right)  ^{\frac{1}{2}}\dot{a}, &
\mbox{when }w=-\frac{2}{3}.
\end{cases}
\label{lapsesolspecial}%
\end{equation}
Finally, the resulting space-time has the line element
\begin{equation}
ds_{w\neq-\frac{2}{3}}^{2}=-\frac{6}{T_{0}a^{2(3w+2)}+\frac{6k}{3w+2}}%
da^{2}+a^{2}\left[  d\chi^{2}+\chi^{2}\mathrm{sinc}^{2}(\sqrt{k}\chi)\left(
d\theta^{2}+\sin^{2}\theta d\phi^{2}\right)  \right]  \label{linelspgen}%
\end{equation}
or
\begin{equation}
ds_{w=-\frac{2}{3}}^{2}=-\frac{6}{T_{0}+6k-12k\ln a}da^{2}+a^{2}\left[
d\chi^{2}+\chi^{2}\mathrm{sinc}^{2}(\sqrt{k}\chi)\left(  d\theta^{2}+\sin
^{2}\theta d\phi^{2}\right)  \right]  \label{linelspgen23}%
\end{equation}
respectively. In this form, where the solution has been obtained in terms of
the lapse, the scale factor $a$ becomes effectively the \textquotedblleft
time\textquotedblright\ variable of the system.

Unfortunately, the generic expression \eqref{parttorsc} giving $T(a)$ for
$w\neq-2/3$, cannot be easily inverted for all values of the parameters. Thus,
it is not trivial to obtain the explicit form of the relative $f(T)$ function
for which the line element \eqref{linelspgen} is the solution. A case for
which this is easily possible, is when $T_{0}=0$. Then, the first of
\eqref{parttorsc} is easily invertible and we obtain from \eqref{partfa}
\begin{equation}
f(T)=\frac{\kappa\rho_{0}}{6^{\frac{3(w+1)}{2}}k}\left(  -\frac{3w+2}%
{k(3w+1)}\right)  ^{\frac{3w+1}{2}}T^{\frac{3(w+1)}{2}}%
\end{equation}
which corresponds to a power-law theory, with the power depending on the
equation of state parameter. Thus, we can infer that the non-zero $T_{0}$
signifies in this case deviations from this particular class of power-law theories.

The situation is easier in the $w=-2/3$ case, where the relative expression in
\eqref{parttorsc} can be inverted to give
\begin{equation}
a=\left[  \frac{6k}{T}W\left(  \frac{e^{\frac{T_{0}}{6k}}}{6k}T\right)
\right]  ^{\frac{1}{2}},
\end{equation}
where $W(z)$ is the Lambert $W$ function which is defined as the solution to
the equation $We^{W}=z$. The corresponding $f(T)$ theory can simply be given
by substituting the above expression in the second branch of \eqref{partfa}. Let us now proceed to see how to quantize this reduced system in each of the two cases.

\subsubsection{The $w\neq-2/3$ case}

In the attempt to construct a quantization scheme for these particular classes
of $f(T)$ theories, we will consider as a given the relations $T=T(a)$ obtained in
\eqref{parttorsc}, and try to write a Lagrangian for the remaining degrees of
freedom. We start by considering the generic case $w\neq-2/3$. We notice that
by crudely substituting the corresponding relations from \eqref{parttorsc} and
\eqref{partfa} into the Lagrangian \eqref{Lag}, we are led to a new, reduced
Lagrangian which reads
\begin{equation}
L_{w\neq-\frac{2}{3}}=\frac{\left(  3\rho_{0}(w+1)a^{-3w}\right)  }{4kN}%
\dot{a}^{2}-N\frac{\rho_{0}(w+1)a^{-3w}\left(  T_{0}(3w+2)a^{2(3w+2)}%
+6k\right)  }{8k(3w+2)}. \label{Laggenw}%
\end{equation}
Interestingly enough, the Euler-Lagrange equations of $L_{w\neq-\frac{2}{3}}$,
\begin{align}
&  N^{2}\left(  T_{0}(3w+2)a^{2(3w+2)}+6k\right)  -6(3w+2)\dot{a}^{2}=0\\
&  2\ddot{a}-3w\frac{\dot{a}^{2}}{a}-2\dot{a}\frac{\dot{N}}{N}+\frac{N^{2}}%
{a}\left(  \frac{3kw}{3w+2}-\frac{T_{0}}{6}(3w+4)a^{6w+4}\right)  =0
\end{align}
give rise to the correct lapse function $N(t)$, as seen in the first of
\eqref{lapsesolspecial}. The latter satisfies both of the above equations.
Hence, we have a Lagrangian giving rise to a dynamically equivalent solution
to that of the gravitational system for the class of theories characterized by
the principal branch of \eqref{partfa}.

Lagrangian \eqref{Laggenw} describes a constrained system, whose Hamiltonian
is (we avoid repeating in detail the formal Dirac procedure since in this case
the treatment is the typical one encountered in most minisuperspace systems)
\begin{equation}
H_{w\neq-\frac{2}{3}}=N\mathcal{H}_{w\neq-\frac{2}{3}}+u_{N}p_{N}%
\end{equation}
where $p_{N}\approx0$ is the primary constraint and
\begin{equation}
\mathcal{H}_{w\neq-\frac{2}{3}}=\frac{ka^{3w}}{3\rho_{0}(1+w)}p_{a}^{2}%
-\frac{\rho_{0}(w+1)}{4}\left[  \frac{3}{(3w+2)a^{3w}}+\frac{T_{0}}%
{2k}a^{3w+4}\right]  \approx0
\end{equation}
is the secondary.

Both of them are first class, so in this case we may proceed in the usual
manner and try to construct a quantum operator for $\mathcal{H}_{w\neq-\frac{2}{3}}$. It is quite convenient in these
cases to exploit the inherent parametrization invariance in these systems to
give to the Hamiltonian constraint a form which will be easier to lead to a
quantum solution. For example, if we adopt the lapse reparametrization
\begin{equation}
N\mapsto\tilde{N}=\frac{\rho_{0}}{\frac{1}{8}(w+1)a^{-3w}\left(  \frac
{T_{0}a^{6w+4}}{k}+\frac{6}{3w+2}\right)  }N,
\end{equation}
then we may write the same Hamiltonian as
\begin{equation}
H_{w\neq-\frac{2}{3}}=\tilde{N}\tilde{\mathcal{H}}_{w\neq-\frac{2}{3}%
}+u_{\tilde{N}}p_{\tilde{N}}%
\end{equation}
where now
\begin{equation}
\tilde{\mathcal{H}}_{w\neq-\frac{2}{3}}=\frac{1}{h(a)}p_{a}^{2}-\rho_{0}%
^{2}\approx0. \label{Hsptild}%
\end{equation}
in which we have
\begin{equation}
h(a)=\frac{3T_{0}(w+1)^{2}}{8k^{2}}a^{4}+\frac{9(w+1)^{2}}{4k(3w+2)a^{6w}}.
\end{equation}
The Hamiltonian constraint \eqref{Hsptild} can be easily assigned to a
Hermitian operator whose action on the wave function will result in a typical
eigenequation. Such an operator is the following
\begin{equation}
\widehat{\tilde{\mathcal{H}}}_{w\neq-\frac{2}{3}}=\frac{1}{\mu(a)}%
\frac{\partial}{\partial a}\left(  \mu(a)\frac{1}{h(a)}\frac{\partial
}{\partial a}\;\right)  -\rho_{0}^{2}, \label{partoperator}%
\end{equation}
which is Hermitian under the measure $\mu(a)$. For the natural measure $\mu(a)=\sqrt{h(a)}$ the differential part of the above operator becomes an one-dimensional
Laplacian \cite{Dimgenquant}. In this case, the solution to $\widehat{\tilde{\mathcal{H}}}%
_{w\neq-\frac{2}{3}}\Psi=0$, yields the wave function
\begin{equation}
\Psi(a)=C_{1}\exp\left(  \mathrm{i}\,\rho_{0}\int\!\!\sqrt{h(a)}da\right)
+C_{2}\exp\left(  -\mathrm{i}\,\rho_{0}\int\!\!\sqrt{h(a)}da\right)  .
\label{parwavefunc}%
\end{equation}
Assuming that we choose either an outgoing or an ingoing wave, i.e. have
either one of the $C_{i}$'s zero, then the probability amplitude, $P$, ends up
to be proportional to
\begin{equation}
P(a)\propto\mu(a)\Psi^{\ast}\Psi=\sqrt{h(a)}. \label{proampl}%
\end{equation}

\begin{figure}[ptb]
{\normalsize
\includegraphics[width=1\textwidth]{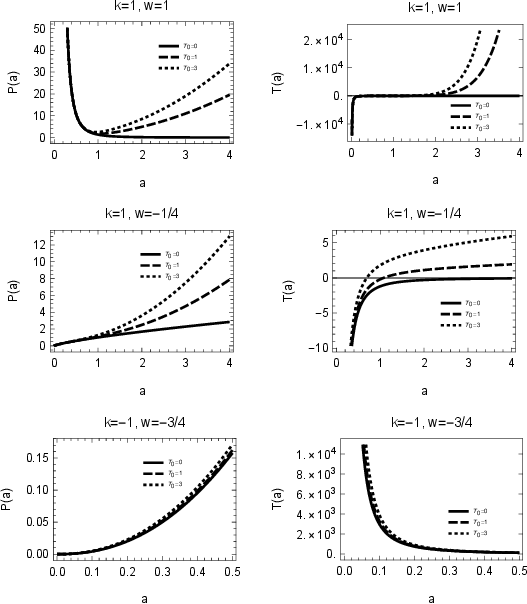} }\caption{A comparison of
the probability amplitude $P$ and the torsion scalar $T$, as functions of the
scale factor $a$, for different $f(T)$ theories characterized by different
values of $T_{0}$.}%
\label{fig1}%
\end{figure}

In Fig. \ref{fig1} we present the graphs of the probability amplitude
\eqref{proampl} and of the torsion scalar (first of Eq. \eqref{parttorsc})
with respect to the scale factor $a$. The latter is the classical relation
which we enforced in order to construct the reduced minisuperspace Lagrangian.
The values of the parameters have been chosen so that the solution
\eqref{linelspgen} represents a Lorentzian line element for all values of $a$.
In general, by inspecting \eqref{linelspgen}, we have the following
possibilities for a Lorentzian metric: i) $T_{0}>0$, $k>0$ and $w>-2/3$, ii)
$T_{0}>0$, $k<0$ and $w<-2/3$, iii) $T_{0}>0$, $k>0$, $w<-2/3$, with an upper
bound on $a(t)$, iv) $T_{0}>0$, $k<0$, $w>-2/3$, with a nonzero lower bound on
$a(t)$, v) $T_{0}<0$, $k>0$, $w>-2/3$ with a fully bounded $a(t)$ and vi)
$T_{0}<0$, $k<0$, $w<-2/3$, again with a totally bounded scale factor. In Fig.
\ref{fig1} we present graphs belonging to cases i) and ii), where the scale
factor can take values in the whole real line. The plots of the (classical) $T(a)$ shows
us the possibly problematic points, where the geometric scalar goes to
infinity. In the first couple of graphs we see a $T(a)$ plot where the torsion
scalar diverges both at $a\rightarrow0$ and at $a\rightarrow+\infty$. At the
same point the probability amplitude seem to also diverge. In the second pair,
there are also two points of divergence - with an exception of the $T_{0}=0$
case - for the $T(a)$. The problematic points are again, $a\rightarrow0$ and
$a\rightarrow+\infty$; the first divergence seems to be \textquotedblleft
cured\textquotedblright\ in the quantum description, in the sense that the
probability amplitude goes to zero at $a=0$. However, for $a\rightarrow
+\infty$ we also see a divergence at $P(a)$. Only the $T_{0}=0$ case is free
of this latter singularity. Finally, the last couple of graphs, shows that in
the hyperbolic case of $k=-1$, the quantum probability amplitude becomes zero
at the point of the classical singularity, which is now only at $a\rightarrow
0$. We need to mention however, that the fact that we partly imposed the
classical solution \eqref{parttorsc}, in order to arrive at a quantum
minisuperspace description, can possibly undermine the effort of lifting the classical
singularity at the quantum level.

As in the previous cases, we can also enforce a crude normalization in terms
of a delta function, by extending the domain of the scale factor $a(t)$ to
negative values as well, and thus consider $a\in R$. The classical solution is in
any case insensitive to the sign of $a(t)$ since only $a(t)^{2}$ appears
inside the metric. With this consideration for the probability integral we
would have, for two states of matter $\rho_{0}$ and $\tilde{\rho}_{0}$,
\begin{equation}%
\begin{split}
|C_{1}|^{2}\int_{-\infty}^{+\infty}\!\!\sqrt{h(a)}\Psi_{\tilde{\rho}_{0}}%
^{*}\Psi_{\rho_{0}}  &  =|C_{1}|^{2}\int_{-\infty}^{+\infty}\!\!\sqrt
{h(a)}e^{-\mathrm{i}\tilde{\rho}_{0}\int\!\!h(a)^{1/2}da}e^{\mathrm{i}\rho
_{0}\int\!\!h(a)^{1/2}da}da\\
&  =|C_{1}|^{2}\int_{-\infty}^{+\infty}\!\!e^{-\mathrm{i}(\tilde{\rho}%
_{0}-\rho_{0})s}ds=2\pi|C_{1}|^{2}\delta(\rho_{0}-\tilde{\rho}_{0}),
\end{split}
\end{equation}
where we made use of the change of variables $a\rightarrow s=\int
\!\!h(a)^{1/2}da$; it can be seen for the various values of the parameters
that when $a\in(-\infty,+\infty)$, then also $s\in(-\infty,+\infty)$.

\begin{figure}[ptb]
{\normalsize
\includegraphics[width=1\textwidth]{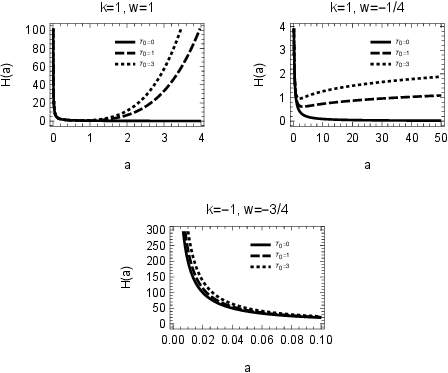} }\caption{The Hubble
function $H(a)$, for different $f(T)$ theories, $k$ and $w$.}%
\label{fig2}%
\end{figure}

For completeness we include the graphs of the Hubble function
\begin{equation}
H(a)=\frac{1}{N}\frac{\dot{a}}{a}=\left(  \frac{1}{6}T_{0}a^{6w+2}+\frac{k}%
{a^{2}(3w+2)}\right)  ^{\frac{1}{2}} \label{Hubexample}%
\end{equation}
with respect to the scale factor $a$, which can be seen in Fig. \ref{fig2}. We
can thus have a sense of the expansion rates that each classical solution implies
for different $f(T)$ theories, characterized by $T_{0}$, in conjunction to the
spatial curvature and the matter content. We observe that the closed
universes, for $T_{0}\neq0$, give rise to ever faster late time acceleration
rates while in the $k=-1$ configuration we get a $H(a)$ function which tends
to zero as $a\rightarrow+\infty$. It is a matter of which term dominates in
Eq. \eqref{Hubexample}.

\subsubsection{The $w=-2/3$ case}

We follow a similar procedure for the $w=-2/3$ case. This time the reduced
minisuperspace Lagrangian that we construct as a function of $N(t)$, $a(t)$ and its
derivative is
\begin{equation}
L_{w=-\frac{2}{3}}=\frac{\rho_{0}a^{2}}{4kN}\dot{a}^{2}-N\rho_{0}a^{2}\left(
\frac{12k\ln a+18k-T_{0}}{24k}-1\right)  .
\end{equation}
Once more, the Euler-Lagrange equations, which are equivalent to
\begin{align}
&  N^{2}-\frac{6\dot{a}^{2}}{-12k\ln a+6k+T_{0}}=0\\
&  \ddot{a}-\frac{\dot{a}\dot{N}}{N}+\frac{\dot{a}^{2}}{a}+\frac{N^{2}}%
{a}\left(  2k\ln a-\frac{T_{0}}{6}\right)  =0,
\end{align}
are satisfied by the gravitational result, i.e. the second of
\eqref{lapsesolspecial}. Thus, we have written a Lagrangian generating the
expected dynamics.

In the exact same manner as before, we write the Hamiltonian and perform a
reparametrization of the lapse function as
\begin{equation}
N\mapsto\tilde{N}=\frac{24k\rho_{0}}{a^{2}(6k+T_{0}-12k\ln a)}N
\end{equation}
to obtain
\begin{equation}
H_{w=-\frac{2}{3}}=\tilde{N}\tilde{\mathcal{H}}_{w=-\frac{2}{3}}+u_{\tilde{N}%
}p_{\tilde{N}}%
\end{equation}
where the Hamiltonian constraint assumes the same form as in \eqref{Hsptild}
\begin{equation}
\tilde{\mathcal{H}}_{w\neq-\frac{2}{3}}=\frac{1}{h(a)}p_{a}^{2}-\rho_{0}%
^{2}\approx0. \label{Hsptild2}%
\end{equation}
where this time the $h(a)$ function is
\begin{equation}
h(a)=\frac{a^{4}(6k+T_{0}-12k\ln a)}{24k^{2}}. \label{ha2}%
\end{equation}
The same analysis as before leads to an operator of the form
\eqref{partoperator} leading to a wave function given by \eqref{parwavefunc},
where of course this time the $h(a)$ function is the one we see in Eq. \eqref{ha2}.

\begin{figure}[ptb]
{\normalsize
\includegraphics[width=1\textwidth]{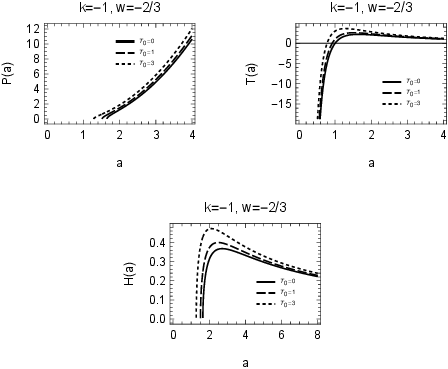} }\caption{The probability
amplitude $P$, the torsion scalar $T$ and Hubble function $H$, as functions of
the scale factor $a$, for $k=-1$, $w=-2/3$ and for different $T_{0}$ values.}%
\label{fig3}%
\end{figure}

In Fig. \ref{fig3} we give once more the probability amplitudes, the torsion
scalar and the Hubble function with relation to the scale factor $a$. We have chosen to depict the $k=-1$ case, which leads to the scale
factor being bound from below to a nonzero value. The geometric singularity of
$T\rightarrow-\infty$ for this value seems to be resolved since it corresponds
to zero probability amplitudes.

The procedure we followed in these last sections is strikingly different from the one which led as to the introduction of fractional operators in section \ref{secfrac}. This is owed to the fact that, in the $k\neq0$ case, we were not able to assign a quantum operator to the obtained Hamiltonian after a pair of canonical variables with respect to Dirac brackets had been introduced. What we did instead here is exploit the fact that we deal with a particular sub-case of the $k\neq0$ case, which leads to a specific relation between the scale factor and the torsion scalar. To unveil this specific relation, we partially integrated the classical system and we afterwards quantized the remaining, reduced system. We would intuitively expect that this second approach we follow, for the particular $k\neq0$ case, is an even cruder approximation of the quantum gravitational system (comparing with the usual quantum cosmology). This is due to the aforementioned successive enforcement of classical relations before quantization. Such an approximation could be possibly applied also to the case of section \ref{secfrac}, but not for a generic $f(T)$ function since some information should be needed to obtain a specific $T(a)$ relation. Unfortunately, the result of this section, is not directly comparable with that of the generic $k=0$ case either. This is because, when setting $k=0$ to the staring relation \eqref{denominatorzero}, we obtain the case a trivial theory $f(T)\sim \sqrt{T}$ in $k=0$, whose Lagrangian is a total derivative.

\section{Conclusions}

\label{sec5}

We performed a detailed analysis of canonical quantum cosmology in $f\left(  T\right)-$gravity. With the help of the Dirac-Bergmann algorithm the constraints of the theory are revealed and classified into first and second class. We use the Dirac brackets as the basic relations with respect to which we proceed to the canonical quantization of the system.

Similarly to what happens in the $f(Q)$-theory \cite{QuantfQus}, in the case of a perfect fluid and for a spatially flat FLRW metric, we arrive in a Hamiltonian where the momentum is raised in a power which is not necessarily an integer number. Thus implying the necessity of putting in use the theory of fractional quantum mechanics. We use some examples in our attempt to interpret the square of the modulus of the wave function in a probabilistic manner, which gets maximized in certain values of the torsion scalar $T$, depending on the theory and its parameters. It is quite positive in the examples that we examined, that probability amplitudes can be constructed that tend to zero as the scalar $T$ diverges. In comparison with classical dynamics we also revealed that the quantum description may even allow for values of $T$ which are classically forbidden.

For the $k\neq0$ case we were not able to treat the generic theory at the quantum level, due to not being able to produce an appropriate mapping from the reduced Hamiltonian to a quantum operator. In the presence of spatial curvature, we restricted our study in the specific case, where Eq. \eqref{denominatorzero} holds, leading to a specific relation $T(a)$. We exploited this relation to reduce even further the classical system and perform a canonical quantization on what remains of it. Here, a typical quantization based purely on first class constraints can be applied. The singularity avoidance, in the sense of $P\rightarrow 0$ as $T\rightarrow \infty$ was not as consistent as in the previous case. We cannot infer if this is an effect of reducing further the system by the application of the classical $T(a)$ relation.

Apart from presenting our findings, we meant this work to also have a pedagogical approach. For this reason, we also made a comparison among the minisuperspace Lagrangians emerging in the modifications of the gravitational trinity. We stressed the idiosyncrasy of the Hamiltonian formulation in the case of teleparallel cosmology. The application of the Dirac-Bergmann algorithm requires a more careful treatment for systems with second class constraints; one cannot just blindly put in use the exact same process reserved for systems with purely first class constraints. The algorithm is a very important tool not only because it offers a way to proceed in a rigorous manner to the Hamiltonian formulation and subsequently to the canonical quantization, but also because it allows us to distinguish the physical degrees of freedom in a given problem.

In the case of field theories the realization of the Dirac-Bergmann algorithm often requires the circumvention of certain difficulties of technical nature \cite{Sund}, and this is something which has recently been underlined for the case of teleparallel theories of gravity \cite{HeisDB} (see more recently \cite{BahaDB} for an account of how surpassing the inherent difficulties in the case of $f(Q)$ theory to offer a formal count of the physical degrees of freedom). In cosmology however, we only have to deal with systems of finite degrees of freedom, where the application of the algorithm is straightforward. The main difficulty here lies in the quantization method and finding an appropriate mapping yielding the corresponding quantum operators.

This work completes a series of studies on quantum cosmology for the modified
theories of gravity described by the scalar fields which form the trinity of
General Relativity. In the future we plan to extend our analysis in the case
of astrophysical objects.

\begin{acknowledgments}
A.P. was supported in part by the National Research Foundation of South Africa
(Grant Numbers 131604). AP thanks the support of Vicerrector\'{\i}a de
Investigaci\'{o}n y Desarrollo Tecnol\'{o}gico (Vridt) at Universidad
Cat\'{o}lica del Norte through N\'{u}cleo de Investigaci\'{o}n Geometr\'{\i}a
Diferencial y Aplicaciones, Resoluci\'{o}n Vridt No - 096/2022.
\end{acknowledgments}

\end{document}